\documentclass[pra,twocolumn,superscriptaddress]{revtex4-1}

\usepackage[utf8]{inputenc}
\usepackage[english]{babel}
\usepackage[T1]{fontenc}

\usepackage{graphicx}
\usepackage{amsmath}
\usepackage{amsfonts}
\usepackage{amssymb}

\usepackage{braket}
\usepackage{makecell}
\usepackage{dsfont}

\usepackage[breaklinks=true,colorlinks]{hyperref}

\newcommand{\omegac}{\omega_\mathrm{c}}
\newcommand{\kB}{k_\mathrm{B}}

\begin{document}

\title{Multiatom Quantum Coherences in Micromasers as Fuel for Thermal and Nonthermal Machines}

\author{Ceren B. Da\u{g}}
\affiliation{Physics Department, University of Michigan, 450 Church St., Ann Arbor, MI 48109-1040, USA}

\author{Wolfgang Niedenzu}
\affiliation{Department of Chemical Physics, Weizmann Institute of Science, Rehovot 7610001, Israel}

\author{\"{O}zg\"{u}r E. M\"{u}stecapl{\i}o\u{g}lu}
\email{omustecap@ku.edu.tr}
\affiliation{Department of Physics, Koc University, 34450 Sar{\i}yer, Istanbul, Turkey}

\author{Gershon Kurizki}
\affiliation{Department of Chemical Physics, Weizmann Institute of Science, Rehovot 7610001, Israel}

\begin{abstract}
  In this paper we address the question: To what extent is the quantum state preparation of multiatom clusters (before they are injected into the microwave cavity) instrumental for determining not only the kind of machine we may operate but also the quantitative bounds of its performance? Figuratively speaking, if the multiatom cluster is the ``crude oil'', the question is: Which preparation of the cluster is the refining process that can deliver a ``gasoline'' with a ``specific octane''? We classify coherences or quantum correlations among the atoms according to their ability to serve as (i) fuel for nonthermal machines corresponding to atomic states whose coherences displace or squeeze the cavity field, as well as cause its heating; and (ii) fuel which is purely ``combustible'', i.e., corresponds to atomic states that only allow for heat and entropy exchange with the field and can energize a proper heat engine. We identify highly promising multiatom states for each kind of fuel and propose viable experimental schemes for their implementation.
\end{abstract}

\maketitle

\section{Introduction}

The maser (microwave amplification by stimulated emission of radiation) was conceived based on thermodynamic considerations~\cite{lamb_laser_1999,schawlow_infrared_1958}. In its micromaser implementation, coherent radiation is generated by inverted two-level (Rydberg) atoms that are randomly injected into a microwave cavity one by one~\cite{walther_one-atom_1996,PhysRevLett.54.551,PhysRevA.34.3077}. For years, the focus of micromaser studies had been on its quantum-electrodynamics features~\cite{varcoe_preparing_2000,rempe_observation_1987,weidinger_trapping_1999,temnov_superradiance_2005,brune_realization_1987,orszag_quantum_1994,dariano_fine_1995,krause_quantum_1986,casagrande_coherently_2002,PhysRevA.46.5913,PhysRevLett.54.551,PhysRevA.34.3077,scullyqoptics}, including its extensions to the cooperative regime of multiatom clusters that are simultaneously present in the cavity~\cite{orszag_quantum_1994,dariano_fine_1995}, until Scully et al.~\cite{scully2002extracting,Scully07022003} revived the interest in the thermodynamics of such devices. They treated the atomic beam as a thermodynamic resource, since randomly injected atoms, which are discarded (traced out) after they exit the cavity, constitute an effective reservoir (bath) for the cavity field mode (in the Markovian approximation).

\par

The surprising finding of Scully et al.~\cite{scully2002extracting,Scully07022003} was that a beam of three-level atoms with coherence between two of its levels may be viewed as a nonthermal, quantum-coherent (``phaseonium'') bath that, given an appropriate phase $\varphi$ of the interlevel coherence, can thermalize the cavity field to a temperature $T_\varphi>T$, where $T$ is the atoms' temperature without coherence. The dramatic consequence of the higher temperature attainable by the cavity field owing to the phaseonium coherence is a transgression of the nominal Carnot efficiency bound in a heat engine, with the cavity field in the role of a working fluid (WF): If the WF undergoes a cycle where it is coupled to the phaseonium bath in one stroke and to a cold bath at temperature $T_\mathrm{c}$ in another, then the efficiency bound of the engine satisfies $\eta\leq 1-T_\mathrm{c}/T_\varphi$, instead of $\eta\leq 1-T_\mathrm{c}/T$. This landmark proposal has triggered a variety of proposals for engine schemes based on nonthermal baths that are capable of ``super-Carnot'' operation~\cite{0295-5075-88-5-50003,PRE89,PhysRevA.81.052121,PhysRevE.73.036122,PhysRevE.84.051122,huang2012effects,abah2014efficiency,rossnagel2014nanoscale,hardal2015superradiant,turkpence2016quantum}, among them engines fuelled by a squeezed nonthermal bath~\cite{rossnagel2014nanoscale}.

\par

Some of us have recently asserted~\cite{niedenzu2015efficiency} that machines fuelled by nonthermal baths may be divided into two categories according to their operation paradigm:
\begin{itemize}
\item \emph{Machines of the first kind} are those fuelled by a nonthermal bath, such as a squeezed-thermal or coherently-displaced thermal bath, that render the WF steady-state \emph{nonpassive}~\cite{pusz1978passive,lenard1978thermodynamical,allahverdyan2004maximal,binder2015quantum,gelbwaser2013work,gelbwaser2014heat,gelbwaser2015thermodynamics}. Such baths change the machine into a \emph{thermo-mechanical engine} that, unlike a heat engine, is fuelled by both mechanical work and heat imparted by the bath to the WF. The Carnot bound may be transgressed in such machines at the expense of work supplied by the bath. However, their efficiency bound cannot be properly compared with the Carnot bound, since the latter is a restriction imposed by the second law on heat~\cite{schwablbook} but not on work imparted by the bath.
\item \emph{Machines of the second kind} are those where the WF is thermalized by the nonthermal bath, as is the case of an engine fuelled by a phaseonium bath. Such a machine is a proper \emph{heat engine} but the ability of the phaseonium bath to thermalize the WF to a temperature $T_\varphi>T$ elevates its Carnot bound above that associated with an incoherent bath at temperature $T$.
\end{itemize}

\par

Intriguingly, in micromaser setups, a beam of multiatom clusters has been shown to thermalize the cavity-field WF in some cases~\cite{0295-5075-88-5-50003,PRE89}, but also coherently displace~\cite{PhysRevA.81.052121,hardal2015superradiant} or squeeze it~\cite{qamar1990generation}. This implies that the WF may receive both work (and thus become nonpassive) and heat from the bath. The cavity field may thus be the key ingredient in machines of the two kinds surveyed above. However, the criteria whereby atoms in a micromaser can fuel machines of either the first or the second kind are generally unknown, notwithstanding several recent results obtained along this line~\cite{0295-5075-88-5-50003,PRE89}.

\par

Here we pose the question: To what extent is the quantum state preparation of multiatom clusters (before they are injected into the cavity) instrumental for determining not only the kind of machine we may operate but also the quantitative bounds of its performance? Figuratively speaking, if the multiatom cluster is the ``crude oil'', the question is: Which preparation of the cluster is the refining process that can deliver a ``gasoline'' with a ``specific octane''?

\par

To answer this question, we first derive (in Sec.~\ref{sec:model}) a master equation that governs the cavity field under the standard assumption of a short interaction of each atom with the cavity field, $g\tau\ll1$, where $g$ is the atom--field coupling strength and $\tau$ is the interaction time. In this regime, the steady-state density matrix of the cavity field may only be a Gaussian state: thermal, coherently-displaced or squeezed~\cite{walls,adesso2007entanglement}. This restriction is shown to imply (Sec.~\ref{sec:Classification}) that two- or three-atom clusters suffice for the preparation of all Gaussian states of the field, thus making larger clusters qualitatively redundant. Secs.~\ref{sec_squeezing} and~\ref{sec:Results} are devoted, respectively, to the classification of coherences or quantum correlations among the atoms that may be associated with (i) fuel for machines of the first kind that correspond to states whose coherences displace or squeeze the cavity field, as well as cause its heating; and (ii) fuel for machines of the second kind which is purely ``combustible'', i.e., corresponds to atomic states that only allow for heat and entropy exchange with the field. In both Sec.~\ref{sec_squeezing} and~\ref{sec:Results} we identify highly promising multiatom states for each kind of fuel and infer the best parameters relevant to the machine operation. In Sec.~\ref{sec:conclusions} we discuss the results and propose viable experimental protocols for their implementation.

\section{\label{sec:model}Model and Effective Master Equation}

\begin{figure}
\centerline{\includegraphics[width=\columnwidth]{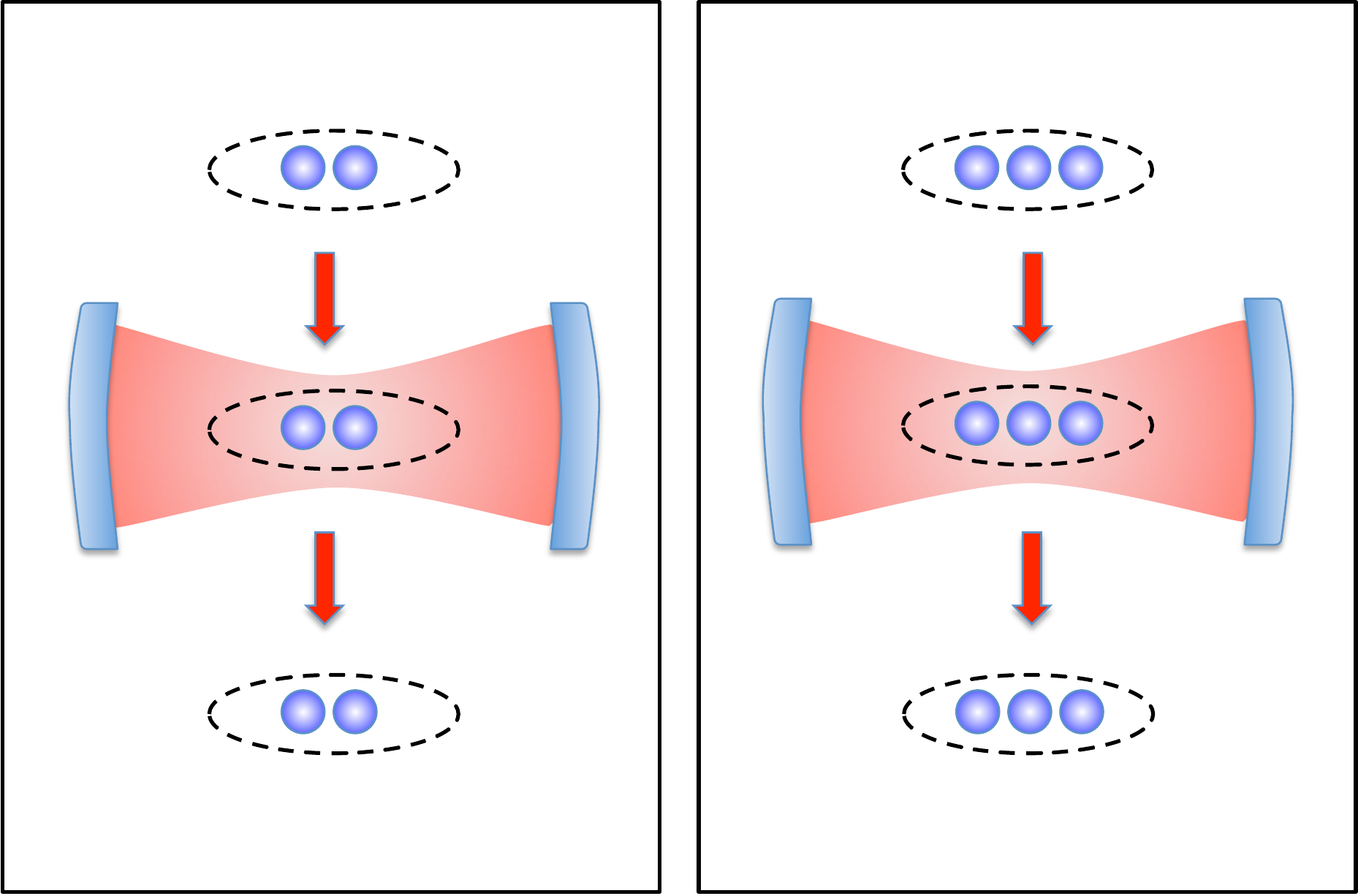}}
\caption{A schematic of the two- and three-atom micromaser model, where clusters of two-level atoms are injected into a single-mode cavity repeatedly in a Poissonian random sequence. The transition time of the atoms through the cavity is much shorter than the cavity lifetime, atomic relaxation and dephasing times or the mean free-time between the interactions, so that there can be at most one cluster present in the cavity at a time. The cavity-mode steady-state crucially depends on the state of the cluster, as shown here.}
\label{sec2fig1}
\end{figure}

We consider a micromaser-type setup wherein the cavity-field mode is the working fluid (WF) that is energized (fuelled) by a beam of two-level atoms which are injected into the cavity at random, Poisson-distributed, times~\cite{PhysRevA.46.5913,PhysRevLett.54.551}. By contrast to the standard micromaser scenario~\cite{PhysRevA.34.3077,MeystreQO}, the atomic beam is here assumed to be composed of $N$-atom clusters that are prepared in a controlled quantum-correlated (entangled) state prior to their injection into the cavity, where they interact with the cavity field simultaneously (see Fig.~\ref{sec2fig1}). Yet, the replacement of single atoms by $N$-atom clusters does not change the basic premise of micromaser theory whereby their random injections allow to treat the atom clusters as an ergodic ``bath'' that is continuously coupled to the cavity mode, so that the latter is governed by a master equation (upon tracing out this ``bath'')~\cite{PhysRevA.46.5913,PhysRevA.34.3077,MeystreQO}. Nevertheless, we will show that quantum coherence or interatomic quantum correlations (entanglement) in the cluster may crucially influence the dynamics of the cavity field. Experimentally feasible schemes for the present scenario will be discussed in Sec.~\ref{sec:conclusions}.

\par

In keeping with the standard assumptions of micromaser theory~\cite{MeystreQO,scullyqoptics}, we take the transition time of the atoms through the cavity to be short enough to neglect atomic relaxation and dephasing as well as cavity loss, and to assume that there can be at most one cluster present in the cavity at a given time. Under these standard assumptions, we may derive a master equation for the dynamics of the cavity field. 

The interaction of the atomic cluster with the cavity is described by the Tavis--Cummings model~\cite{PhysRev.170.379}
\begin{equation}
  H_{\mathrm{TC}} = H_\mathrm{a} + H_\mathrm{c} + H_{\text{int}},
\end{equation}
where the atom, cavity and interaction Hamiltonians are respectively given by
\begin{subequations}
  \begin{align}
    H_\mathrm{a}& = \frac{\hbar \omega_\mathrm{a}}{2} \sum_{j=1}^3 \sigma_j^z; \\
    H_\mathrm{c} &= \hbar \omegac a^{\dagger} a; \\
    H_{\text{int}} &= \hbar g \sum_{j=1}^N ( a \sigma_j^+ + a^{\dagger} \sigma_j^-).\label{eq_H_int}
  \end{align}
\end{subequations}
Here $a,a^{\dagger}$ are the annihilation and creation operators for the cavity field and $\sigma^z_j, \sigma^+_j,\sigma^-_j$ are the $z$, raising and lowering Pauli operators for the $j$th atom with $j=1,\dots,N$. The atomic transition frequency $\omega_\mathrm{a}$ is resonant with the cavity frequency $\omegac$. The interaction between the atoms and the cavity is assumed to be spatially homogeneous with strength $g$.

Under the foregoing assumptions, the combined system of the atomic cluster and the cavity field evolves unitarily during the short interaction time $\tau$. The unitary propagator $U(\tau)=\exp(-iH_\mathrm{int}\tau)$ in the interaction picture can be analytically computed to second order in $g\tau$ (see Appendices~\ref{app_one_atom}--\ref{app_three_atoms}). Denoting the injection time of the $j$th cluster into the cavity by $t_j$, the evolution of the reduced density operator of the field mode, which is obtained upon tracing out the atoms, reads~\cite{schaller,gardiner}
\begin{equation}
\rho (t_j + \tau) = \operatorname{Tr}_{a}[U(\tau) \rho_{a} \otimes \rho (t_j) U^{\dagger}(\tau)] 
\equiv S(\tau) \rho(t_j).
\label{sec2eq4}
\end{equation}
Here $\rho_{a}$ is the initial density operator of the atomic cluster and $S(\tau)$ is a superoperator that propagates the cavity state $\rho(t_j)$ to $\rho (t_j + \tau)$. The atomic clusters arrive randomly at a rate $p$ and pass through the cavity within a time interval of $(t,t+\delta t)$ with a probability of $p \delta t$. The field changes according to $S(\tau)$ when a cluster is present and otherwise does not change at all, so that the overall change of the field state is
\begin{equation}
\rho(t+\delta t) = p \delta t S(\tau) \rho(t) + (1-p\delta t) \rho(t).	\label{sec2eq5}
\end{equation} 
For $\delta t \rightarrow 0$ we obtain the master equation
\begin{equation}
\dot{\rho}(t) = p \left[ S(\tau) - 1 \right] \rho(t) \label{sec2eq6}
\end{equation}
which describes the Markovian dynamics of the single-mode cavity~\cite{PhysRevA.81.052121,PRE89,PhysRevE.73.036122,PhysRevA.34.3077}. Here we do not include the usual cavity decay term, in order to clearly identify the role of coherences for the field evolution, particularly whether it thermalizes or not, but it is straightforward to do so.

\begin{figure}
\centerline{\includegraphics[width=\columnwidth]{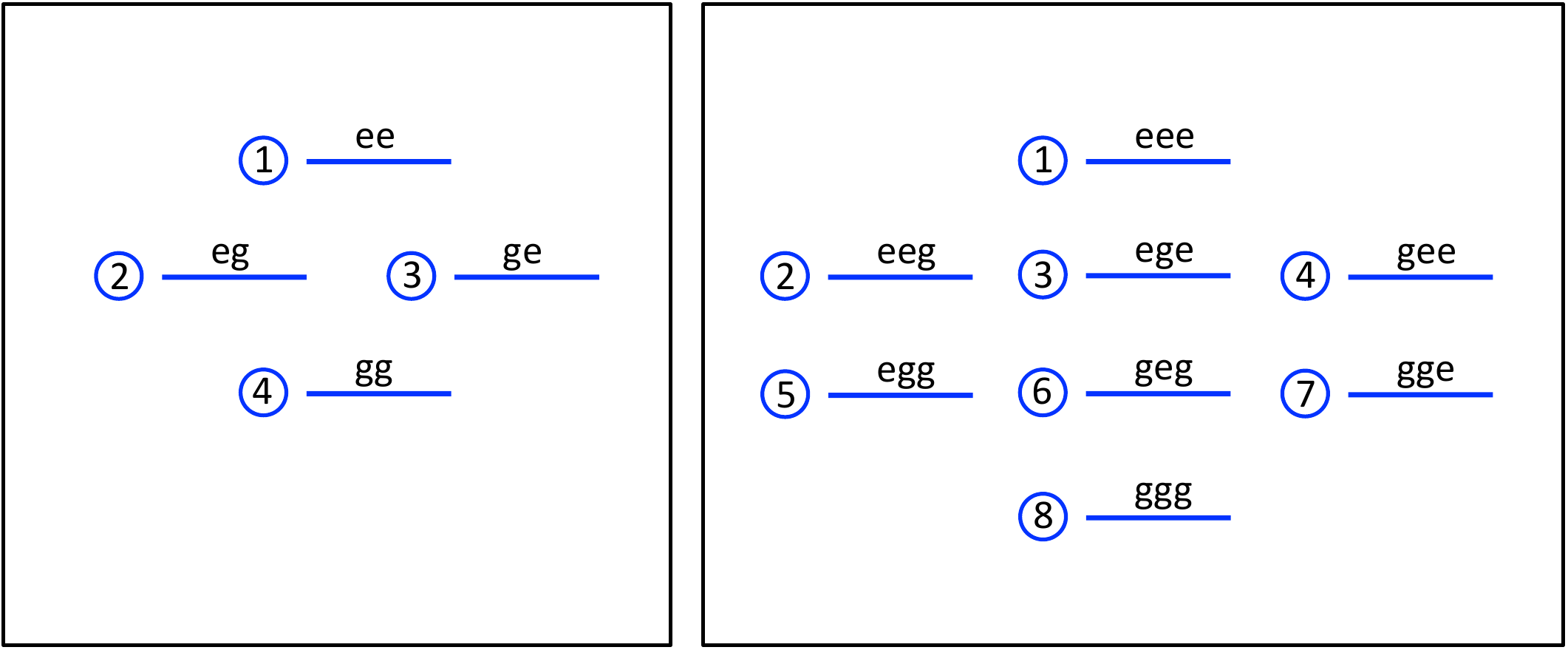}}
\caption{Energy levels of clusters of two- and three two level atoms. The numbers next to the levels correspond to the indices used in the text 
to denote their corresponding position in the natural basis.}
\label{sec2fig2}
\end{figure}

The master equation~\eqref{sec2eq6} can be rewritten as
\begin{equation}
\dot{\rho}(t) = p \left[ \sum_{i,j=1}^N a_{ij} \sum_{n=1}^N U_{ni}(\tau) \rho(t) [U_{nj}(\tau)]^{\dagger}- \rho(t) \right], \label{sec2eq9}
\end{equation}
where $a_{ij}$ denote the matrix elements of $\rho_\mathrm{a}$. In the standard basis of energy-state products the diagonal elements $a_{ii}$ are the populations and the off-diagonal elements $a_{ij}$ with $i\neq j$ are coherences or correlations, respectively. This standard basis is $\{\Ket{e},\Ket{g}\}$ for one atom and shown in Fig.~\ref{sec2fig2} for two ($\{\Ket{ee},\Ket{eg},\Ket{ge},\Ket{gg}\}$) and three atoms, respectively.

Using the explicit forms of the respective propagators $U(\tau)$ for one-, two- and three-atom clusters [Eqs.~\eqref{eq_app_Utau_one},~\eqref{eq_app_Utau_two} and~\eqref{eq_app_Utau_three} in the Appendix],
the master equation~\eqref{sec2eq9} can be expressed in the illuminating form
\begin{equation}
\dot{\rho} \approx -i \left[H_{\text{eff}},\rho \right] + \mathbb{L}_s \rho + \mathbb{L} \rho.
\label{sec:single:eq1}
\end{equation}
Here the first term corresponds to the effect of a coherent drive applied to the cavity, which is described by the effective Hamiltonian 
\begin{equation}
H_{\text{eff}} = p g \tau \left(\lambda a^{\dagger} + \lambda^* a\right). \label{sec:single:eq2}
\end{equation}
The Lindbladian $\mathbb{L}_s$ in Eq.~\eqref{sec:single:eq1} describes a squeezing process and is given by
\begin{equation}
\mathbb{L}_s \rho =  \mu\left(\xi \mathbb{L}^e_s + \xi^* \mathbb{L}^d_s \right),
\label{sec:single:eq3}
\end{equation}
where $\mu = p(g \tau)^2$ is an effective coupling rate. The squeezing excitation and de-excitation Lindbladians are
\begin{subequations}
  \begin{align}
    \mathbb{L}^e_s &= 2 a^{\dagger} \rho a^{\dagger}  - a^{\dagger} a^{\dagger} \rho- \rho a^{\dagger} a^{\dagger}\\
    \mathbb{L}^d_s &= 2 a \rho a - a a \rho- \rho a a,
  \end{align}
\end{subequations}
respectively \cite{walls}.
The Lindbladian $\mathbb{L}$ is given by
\begin{equation}
\mathbb{L} \rho = \mu\left(\frac{r_e}{2} \mathbb{L}_e + \frac{r_g}{2} \mathbb{L}_d \right),
\label{sec:single:eq4}
\end{equation}
where
\begin{subequations}
  \begin{align}
    \mathbb{L}_d &= 2 a \rho a^{\dagger} - a^{\dagger} a \rho - \rho  a^{\dagger} a\\
    \mathbb{L}_e &= 2 a^{\dagger} \rho a - a a^{\dagger} \rho - \rho a a^{\dagger}
  \end{align}
\end{subequations}
are the Lindbladians for incoherent de-excitation and excitation, respectively. The coefficients for different cluster sizes are shown in Table~\ref{table_coefficients}.

\begin{table*}
  \centering
  \begin{tabular}{|c|c|c|c|}
    \hline
    & 1 atom & 2 atoms & 3 atoms \\
    \hline
    $r_e$ & $a_{11}$ & $2a_{11}+a_{22}+a_{33}+a_{23}+a_{32}$ & $3a_{11} + 2D_E + D_W + C_E + C_W$ \\
    \hline
    $r_g$ & $a_{22}$ & $2a_{44}+a_{22}+a_{33}+a_{23}+a_{32}$ & $3a_{88} + 2D_W + D_E + C_E + C_W$ \\
    \hline
    $\lambda$ & $a_{12}$ & $a_{12}+a_{13}+a_{24}+a_{34}$ & \makecell{$a_{25} + a_{35}  + a_{46} + a_{47} + a_{26} + a_{37}$ \\ $+ a_{12} + a_{13} + a_{14} + a_{58} + a_{68} +  a_{78}$} \\
    \hline
    $\xi$ & $0$ & $a_{14}$ & $a_{28} +  a_{38} + a_{48} +a_{15} + a_{16} + a_{17}$ \\
    \hline
  \end{tabular}
  \caption{Coefficients of the master equation~\eqref{sec:single:eq1} for different cluster sizes. For later convenience we have defined for three-atom clusters the abbreviations $D_E = a_{22} + a_{33} + a_{44}$, $D_W = a_{55} + a_{66} + a_{77}$, $C_E = a_{23} + a_{24} + a_{32} + a_{34} + a_{42} + a_{43}$ and $C_W = a_{56} + a_{65} + a_{57} + a_{75} + a_{67} + a_{76}$.}\label{table_coefficients}
\end{table*}

The master equation~\eqref{sec:single:eq1} allows for the generation of arbitrary (Gaussian) field states, i.e., thermal-, displaced- and squeezed states. Higher-order (i.e., non-Gaussian) processes cannot be induced by a second-order master equation. The case $N=2$ is the minimal cluster generating these processes: Adding another particle ($N=3$) does not make a qualitative difference compared to Eq.~\eqref{sec:single:eq1}.

\section{\label{sec:Classification}Classification of Coherences as Different Types of Fuel}

\begin{figure}
\centerline{\includegraphics[width=\columnwidth]{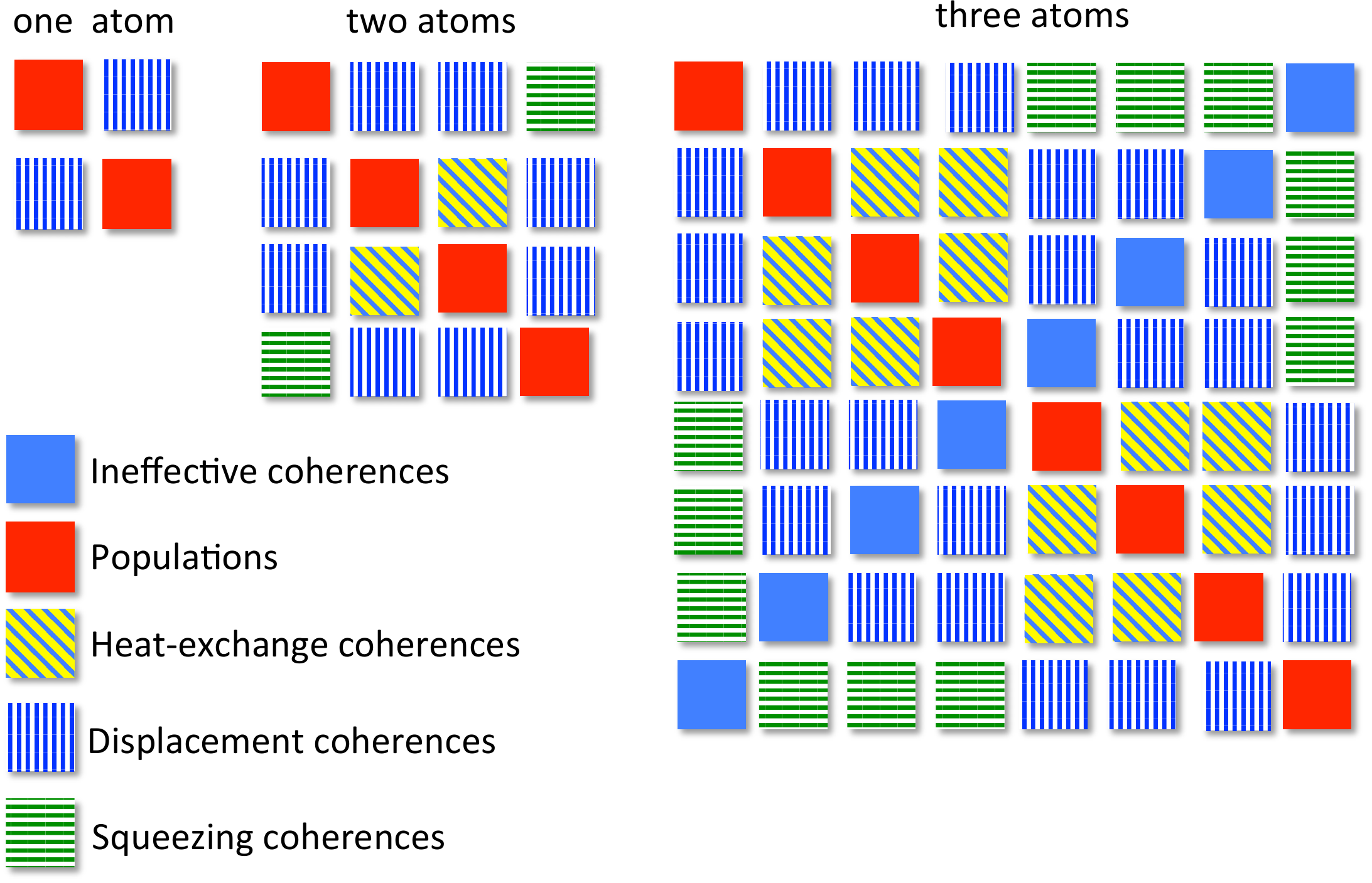}}
\caption{Density matrix of the atomic cluster for one (left), two (middle) and three (right) atoms, respectively, with color- and pattern-filled squares representing the different roles of the coherences with respect to the cavity-field evolution described by the master equation~\eqref{sec:single:eq1}. Red plain dark squares are populations and light blue squares are ineffective coherences. Yellow diagonal striped squares are zeroth order coherences that can contribute to thermalization. Dark
blue vertical striped squares are first order coherences that can contribute to the coherent displacement of the cavity field. Green horizontal striped squares
are second order coherences contributing to the squeezing of the cavity field.}
\label{sec2fig4}
\end{figure}

The key observation we infer from Eq.~\eqref{sec:single:eq1} is that coherences or correlations in the multiatom cluster may be classified according to the disjoint blocks in the density matrix $\rho_\mathrm{a}$ that are associated with qualitatively different terms in the master equation, each term giving rise to a different kind of field dynamics. Figuratively, the different coherences are different types of fuel for the cavity-field WF (Fig.~\ref{sec2fig4}): 
\begin{itemize}
\item The blocks adjacent to the main diagonal of the $\rho_\mathrm{a}$ matrix in the standard basis of Fig.~\ref{sec2fig2} contain coherences that can only induce absorption- and emission processes in the field (WF), as they are associated to $\mathbb{L}\rho$ in the master equation~\eqref{sec:single:eq1}. We shall refer to these elements as \emph{heat-exchange coherences}. They have a caloric (``flammable'') value, i.e., they may contribute to the thermalization of the cavity field. Heat-exchange coherences do not arise in the single-atom case, as they correlate states of the same energy, e.g., $\Ket{eg}\!\Bra{ge}$ in two-atom clusters and $\Ket{eeg}\!\Bra{ege}$ in three-atom clusters.
\item \emph{Displacement coherences} associated with the $-i[H_\mathrm{eff},\rho]$ term in the master equation~\eqref{sec:single:eq1} arise for all cluster sizes as they correlate states differing by one excitation, i.e., $\Ket{e}\!\Bra{g}$ and $\Ket{g}\!\Bra{e}$ in single atoms, $\Ket{eg}\!\Bra{ge}$ and its Hermitian conjugate in two-atom clusters and, say, $\Ket{eeg}\!\Bra{geg}$ in three-atom clusters.
\item \emph{Squeezing coherences} correspond to an exchange of two excitations and may exist in two-atom clusters in the form of $\Ket{ee}\!\Bra{gg}$ and its Hermitian conjugate, or in three-atom clusters in, say, the form $\Ket{eeg}\!\Bra{ggg}$. 
\end{itemize}
Those matrix elements of $\rho_\mathrm{a}$ that do not contribute to the field evolution shall be called \emph{ineffective coherences}.

\par

\begin{figure}
\centerline{\includegraphics[width=\columnwidth]{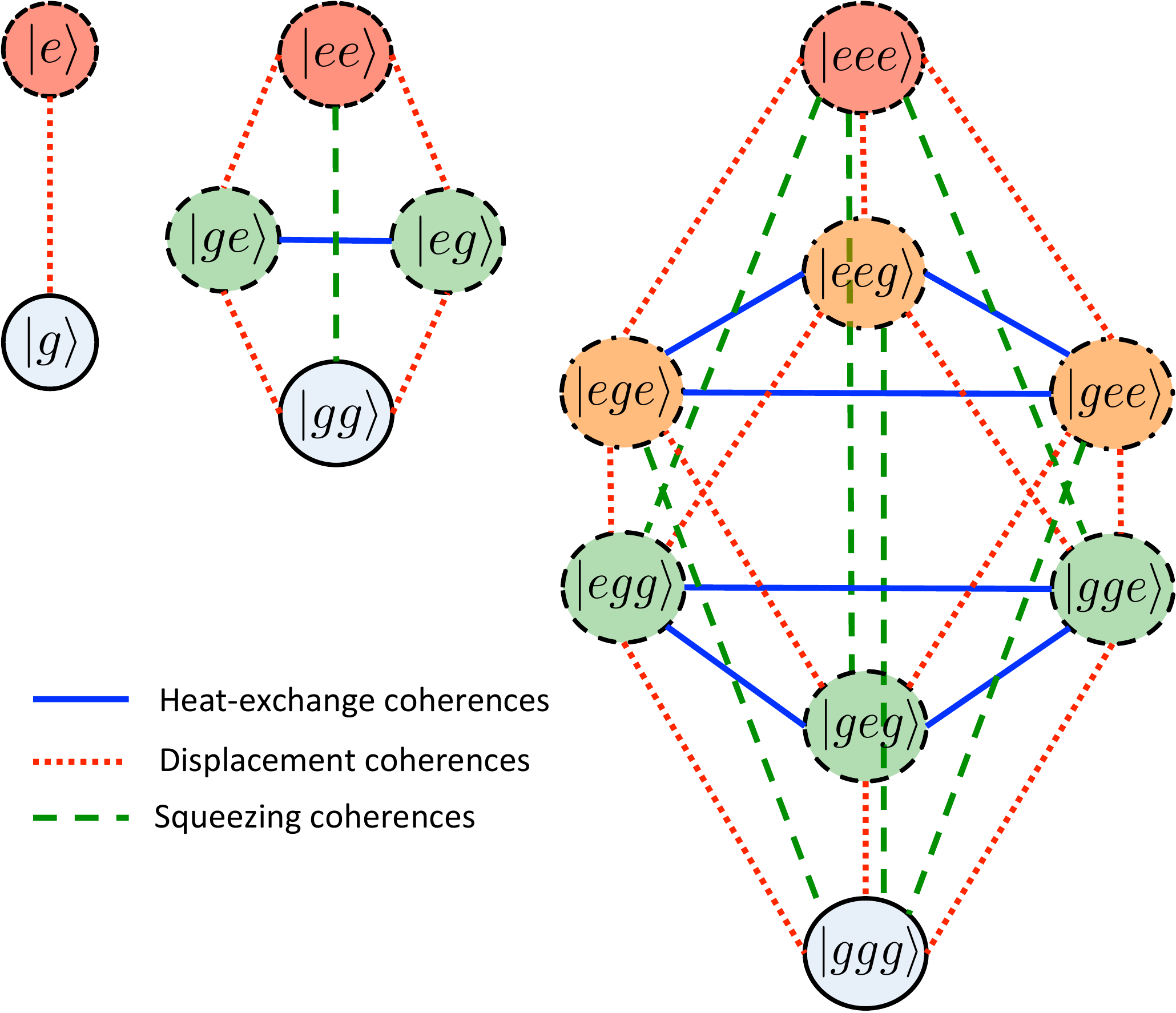}}
\caption{Trees of coherence for one- (left), two- (middle) and three- (right) atom clusters, respectively. The circles are the basis states, where same colour indicates the same number of excitations. Blue solid lines indicate heat-exchange coherences that may contribute to thermalization of the cavity field. These coherences between states with the same number of excitations only appear in the multipartite case. Displacement coherences (red dotted lines) between states differing by one excitation arise for all particle numbers. Squeezing coherences (green dashed lines) are between states differing by two excitations.}
\label{sec2fig3}
\end{figure}

Larger cluster sizes will not change the qualitative features of the master equation~\eqref{sec:single:eq1}, which holds to second order in $g\tau$ and thus may only induce the same second-order (Gaussian) processes as listed above. The different types of coherences and their relation to the number of excitations is illustrated in the tree diagram of Fig.~\ref{sec2fig3} (which should not be confused with the same term in graph theory).

\section{Correlated Atomic Clusters as Fuel for Machines of the First and Second Kind}\label{sec_fuel}

The beam of atomic clusters interacting with a cavity mode can realize one of two operation paradigms~\cite{niedenzu2015efficiency}:
\begin{itemize}
\item If displacing or squeezing coherences are present in the bath, the cavity state becomes \emph{nonpassive} (displaced or squeezed, respectively), which implies that not only heat but also work has been transferred from the bath to the cavity mode. Consequently, a machine fuelled by such a bath is a \emph{machine of the first kind} that operates \emph{thermo-mechanically}.
\item If the atomic state \emph{only} contains heat-exchange coherences, the mode is \emph{thermalized} by the bath and only heat is exchanged. Such a setup is thus a viable implementation of a \emph{heat engine} powered by a nonthermal bath, which has been dubbed a \emph{machine of the second kind}.
\end{itemize}

\subsection{Conditions for fuelling machines of the first kind}\label{sec_squeezing}

Under what conditions does the master equation~\eqref{sec:single:eq1} possess a nonpassive (nonthermal) steady state of the cavity mode that is required for a machine of the first kind (a thermo-mechanical machine)? What is the nature of such a state?

\par

To obtain an insight into these questions, we write the (Ehrenfest) equations of motions of the field mean value, variance and mean intensity (photon number) in terms of the $\rho_\mathrm{a}$ matrix elements that are grouped into the coefficients $r_g$, $r_e$, $\lambda$ and $\xi$ as detailed in Table~\ref{table_coefficients},
\begin{subequations}
  \begin{align}
\Braket{\dot{a}(t)} &= -\frac{\mu}{2} (r_g-r_e) \Braket{a(t)} - i p g \tau \lambda, \label{sec:single:eq5}\\
\Braket{\dot{a}^2(t)} &= -\mu (r_g-r_e)\Braket{a^2(t)} - 2i p g \tau \lambda \Braket{a(t)}- 2\mu\xi,\label{sec:single:eq6}\\
\Braket{\dot{n}(t)} &= -\mu (r_g-r_e) \Braket{n(t)} + \mu r_e\notag\\&\quad- i p g \tau( \lambda \Braket{a^{\dagger}(t)} -  \lambda^* \Braket{a(t)}),
\label{sec:single:eq7}
\end{align}
\end{subequations}
whose steady-state solutions read 
\begin{subequations}
  \begin{align}
\Braket{a}_{\text{ss}} &= \Braket{a^{\dagger}}_{\text{ss}}^* = - \frac{2i \lambda}{g\tau (r_g-r_e)},
\label{sec:single:eq8}\\
\Braket{a^2}_{\text{ss}} &= \Braket{\left(a^{\dagger}\right)^2}_{\text{ss}}^* = - 2 \left(\frac{\xi}{r_g-r_e}+\frac{2 \lambda^2}{(g\tau)^2(r_e-r_g)^2}\right),
\label{sec:single:eq10}\\
\Braket{n}_{\text{ss}} &= \frac{r_e}{r_g-r_e} + \frac{4 |\lambda|^2}{(g \tau)^2 (r_g-r_e)^2}.\label{sec:single:eq9}
\end{align}
\end{subequations}
We see that a nonzero $\lambda$ increases the thermal mean photon number by coherently displacing the cavity field to a nonzero expectation value [Eqs.\eqref{sec:single:eq8} and~\eqref{sec:single:eq9}]. Accordingly, the cavity field attains thermal-coherent character. By contrast, a nonzero $\xi$ introduces quadrature squeezing to the cavity field and hence the cavity field acquires a squeezed-thermal character.

\par

\begin{figure}
  \centering
  \includegraphics[width=0.9\columnwidth]{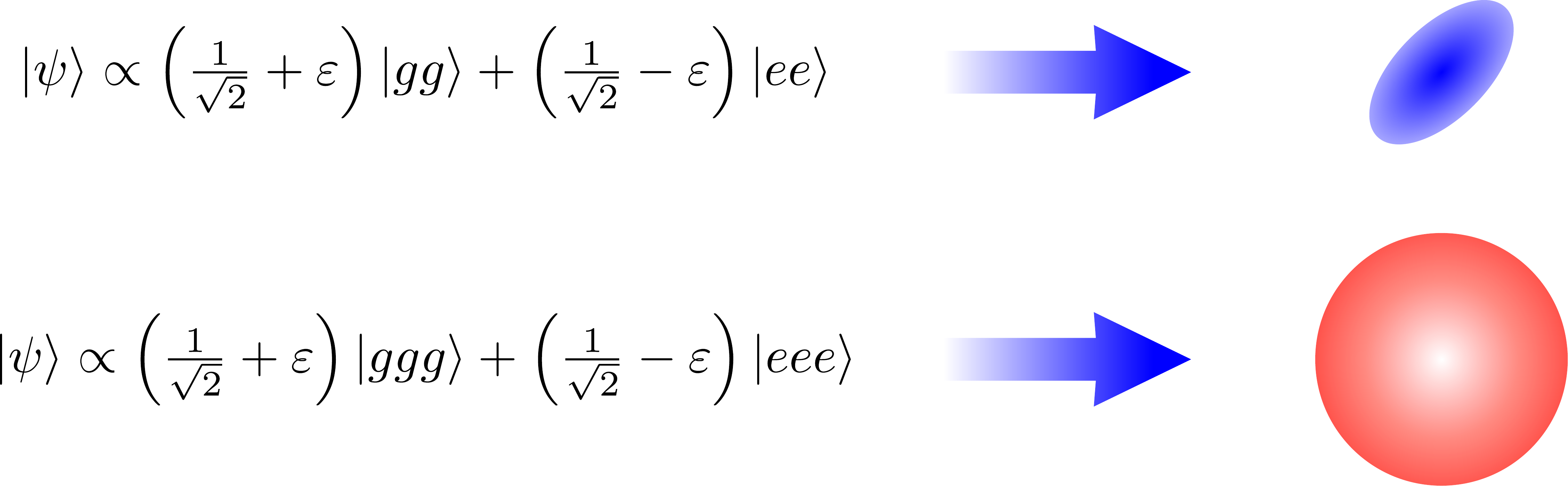}
  \caption{A doubly-excited state (top) gives rise to a squeezed state of the cavity field. By contrast, a triply-excited state (bottom) thermalizes the cavity to an ultrahigh temperature.}\label{fig_states}
\end{figure}

\par

\begin{figure}
  \centering
  \includegraphics[width=0.9\columnwidth]{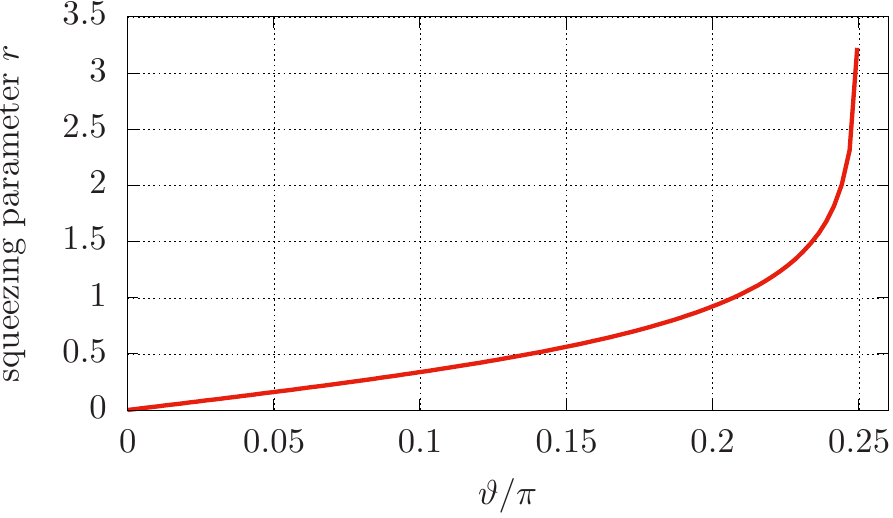}
  \caption{Squeezing parameter~\eqref{eq_master_squeezing_two_particles_r} as a function of the coefficient $\vartheta<\pi/4$ of the two-atom state~\eqref{eq_psi_squeezing}.}\label{fig_squeezing}
\end{figure}

As an example, consider the two-atom state
\begin{equation}\label{eq_psi_squeezing}
  \Ket\psi=\cos\vartheta\Ket{gg}+\sin\vartheta\Ket{ee}.
\end{equation}
For $\sin^2\vartheta<\frac{1}{2}$ (for simplicity we restrict the angle to $0\leq\vartheta<\pi/4$) this state gives rise, according to Eq.~\eqref{sec:single:eq1}, to the master equation (cf. Fig.~\ref{fig_states})
\begin{equation}\label{eq_master_squeezing_two_particles}
  \dot\rho=\mu\frac{1}{2}\sin(2\vartheta)\mathbb{L}_s\rho+\mu\left[\cos^2\vartheta\,\mathbb{L}_g+\sin^2\vartheta\,\mathbb{L}_e\right],
\end{equation}
which may be cast into the standard form that yields thermal-squeezed solutions~\cite{walls,breuerbook}
\begin{equation}\label{eq_master_squeezing_standard}
  \dot\rho=\kappa M \mathbb{L}_s\rho+\kappa\left[(N+1)\mathbb{L}_g+N\mathbb{L}_e\right],
\end{equation}
with the coefficients
\begin{subequations}\label{eq_master_squeezing_standard_coefficients}
  \begin{align}
    N&=\bar{n}(\cosh^2r+\sinh^2r)+\sinh^2r\\
    M&=\cosh r\sinh r (2\bar{n}+1).
  \end{align}
\end{subequations}
Here $r$ denotes the squeezing parameter and $\bar{n}$ the ambient photon number. Upon comparing Eqs.~\eqref{eq_master_squeezing_two_particles},~\eqref{eq_master_squeezing_standard} and~\eqref{eq_master_squeezing_standard_coefficients} we find
\begin{subequations}
  \begin{align}
    \kappa&=\mu\cos(2\vartheta)\\
    \bar{n}&=0\\
    r&=\operatorname{atanh}\left(\tan\vartheta\right).\label{eq_master_squeezing_two_particles_r}
  \end{align}
\end{subequations}
In conventional experimental squeezing schemes, squeezing parameters range from $r\sim0.4$~\cite{wenger2004nongaussian,su2007experimental} up to $r\approx 1.46$~\cite{eberle2010quantum}. Remarkably, the squeezing parameter~\eqref{eq_master_squeezing_two_particles_r} may greatly surpass existing values if we choose $\vartheta\rightarrow \pi/4$ (see Fig.~\ref{fig_squeezing}).

\par

Let us note that the existence of a steady-state solution (whether coherent or not) requires $r_g>r_e$, implying $\sin^2\vartheta<\frac{1}{2}$. We will discuss this point in more detail in what follows.

\subsection{\label{sec:Results}{Conditions for fuelling machines of the second kind}}

It is thus clear that thermal equilibrium of the cavity field requires $\lambda=\xi=0$ so that all coherent processes vanish. This can be achieved by either setting the displacement- and squeezing coherences to zero in Fig.~\ref{sec2fig4} or making their respective contributions to the coefficients $\lambda$ and $\xi$ (see Table~\ref{table_coefficients}) to cancel each other. Under these conditions that are required for thermalization of the cavity field, the master equation reads
\begin{equation}\label{eq_master_thermal}
  \dot{\rho}=\frac{\mu r_g}{2}\mathbb{L}_g\rho+\frac{\mu r_e}{2}\mathbb{L}_e\rho,
\end{equation}
where $\mu r_g/2$ is the rate of emission of quanta into the bath and $\mu r_e/2$ is the absorption rate of quanta from the bath. By virtue of the Kubo--Martin--Schwinger (KMS) detailed-balance condition~\cite{breuerbook}, it is then possible to attribute a temperature $T$ to the effective bath through
\begin{equation}\label{eq_kms}
  r_e = \exp\left(-\frac{\hbar\omegac}{\kB T}\right) r_g.
\end{equation}
This temperature $T$ is only positive and finite if $r_g>r_e$. 

\par

Let us pretend that this effective bath is composed of fictitious oscillators. Then Eq.~\eqref{eq_kms} yields
\begin{equation}\label{eq_kms_2}
  \frac{r_e}{r_g}=\frac{\bar{n}}{\bar{n}+1},
\end{equation}
where 
\begin{equation}
  \bar{n}:=\frac{1}{\exp\left(\frac{\hbar\omegac}{\kB T}\right)-1}
\end{equation}
is the effective thermal excitation of the bath (mean number of fictitious quanta) at the cavity frequency $\omegac$. The denominator in Eq.~\eqref{eq_kms_2} then corresponds to the sum of stimulated and spontaneous emission into the effective heat bath.

\par

This effective description of the effective bath is vindicated by the steady-state solution of the master equation~\eqref{eq_master_thermal} which is the thermal (Gibbs) state~\cite{breuerbook}
\begin{equation}
  \rho_\mathrm{ss}=\frac{1}{Z}\exp\left(-\frac{1}{\kB T} H_\mathrm{c}\right)
\end{equation}
of the cavity mode, where $Z$ denotes the partition function. Due to the unbounded character of the Hamiltonian $H_\mathrm{c}$, this steady-state solution only exists if $0\leq T<\infty$, i.e., if $r_g>r_e$. The case $r_g=r_e$, formally resulting in an infinite bath temperature, can be identified as the maser threshold~\cite{scully_quantum_1967} (see also Appendix~\ref{app_maser_threshold}).

We have thus arrived at an important conclusion: A nonthermal beam of atoms interacting with a single cavity mode may act as an effective heat bath for the latter, thereby thermalizing it to a finite temperature $T$, although the quantum state of the atoms may be distinctly \emph{nonthermal}, i.e., the \emph{atomic-cluster state is not associated with the notion of temperature}. Nevertheless, it will drive the cavity field mode into a Gibbs state with a finite and positive temperature provided the cavity mode is below the maser threshold. This conclusion is consistent with the well-known fact that the regime below the micromaser threshold is thermal radiation with a thermodynamic equilibrium temperature~\cite{scully_quantum_1967,davidovich_sub-poissonian_1996}. Here, however, this temperature $T$ of the cavity field depends explicitly on the coherences and correlations of the atoms.

\par

In Table~\ref{table_values} we present the explicit dependence of the temperature $T$, the steady-state photon number $\langle n\rangle_\mathrm{ss}$, and the micromaser threshold on the multiatomic density-matrix parameters from Table~\ref{table_coefficients}. 

\par

In what follows, we focus on the relation between the correlations in distinctly entangled states of the cluster on the temperature and threshold conditions.

\begin{table*}
  \centering
  \begin{tabular}{|c|c|c|c|}
    \hline
    & 1 atom & 2 atoms & 3 atoms \\
    \hline
    \rule{0pt}{3.5ex}$\frac{\kB T}{\hbar\omega_\mathrm{c}}=\ln\left[\left(\frac{r_g}{r_e}\right)\right]^{-1}$ & $\left[\ln{\left(\frac{a_{22}}{a_{11}}\right)}\right]^{-1}$ & $\left[\ln{\left(\frac{2a_{44}+a_{22}+a_{33}+a_{23}+a_{32}}{2a_{11}+a_{22}+a_{33}+a_{23}+a_{32}}\right)}\right]^{-1}$ & $\left[\ln{\left(\frac{3a_{88}+2D_W+D_E+C}{3a_{11}+2D_E+D_W+C}\right)}\right]^{-1}$ \\[1.5ex]
    \hline
    \rule{0pt}{3.5ex}$\langle n\rangle_\mathrm{ss}=\frac{r_g}{r_g-r_e}$ & $\frac{a_{22}}{a_{22}-a_{11}}$ & $\frac{2a_{44}+a_{22}+a_{33}+a_{23}+a_{32}}{2(a_{44}-a_{11})}$ & $\frac{3a_{11} + 2D_E + D_W + C}{3(a_{88} -a_{11})+D_W-D_E}$ \\[1.5ex]
    \hline
    \rule{0pt}{3.5ex}valid for ($r_g>r_e$) & $a_{22}>a_{11}$ & $a_{44}>a_{11}$ & $3a_{88}+D_W>3a_{11}+D_E$ \\[1.5ex]
    \hline
  \end{tabular}
  \caption{Steady-state properties following from the thermal master equation~\eqref{eq_master_thermal} for different cluster sizes. Here we have defined $C=C_E+C_W$.}\label{table_values}
\end{table*}

\subsubsection{\label{sec:wstate}Cavity thermalization via singly-excited entangled three-atom states}

\begin{figure}
\centerline{\includegraphics[width=0.9\columnwidth]{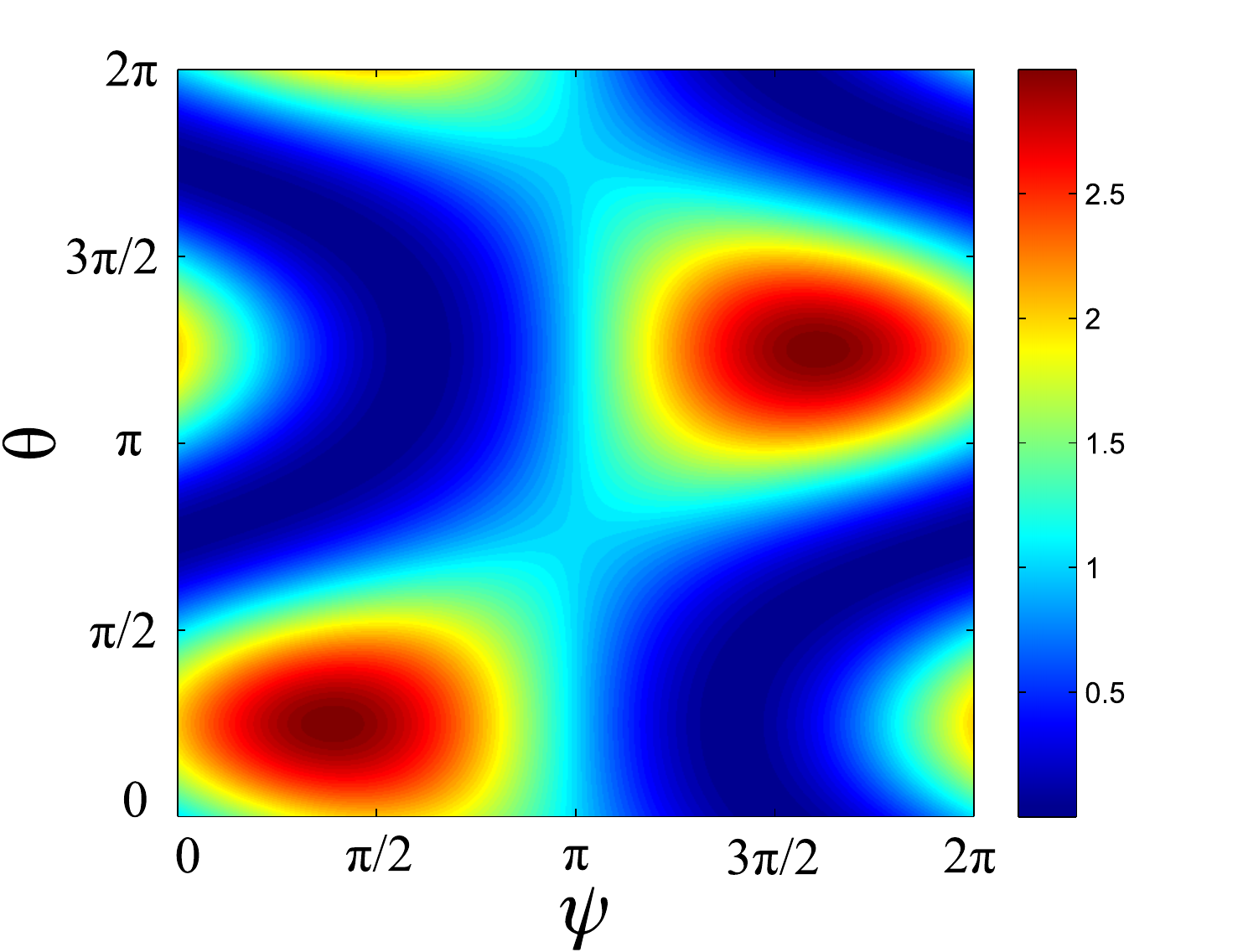}}
\caption{Steady state mean number of photons in a cavity pumped randomly
with three atom clusters in W class states, parameterized with angular variables $\theta$ and $\psi$, when $\delta = 0$ and $\phi = 0$. The symmetric W state yields the largest 
mean photon number in equilibrium and hence can be imagined as the "hottest" effective three atom reservoir among the W class states.}
\label{sec3.1fig1}
\end{figure}

We may parameterize the singly-exited entangled states of three atoms via
\begin{multline}
\Ket{W}_\mathrm{gen} = \cos \theta \cos \frac{\psi}{2} \Ket{egg} + \sin \theta \cos \frac{\psi}{2} \text{e}^{i\phi} \Ket{geg} \\+ \sin \frac{\psi}{2} \text{e}^{i \delta} \Ket{gge}.
\label{sec3.1eq28}
\end{multline}
The mean photon number in terms of the angles then reads (cf. Table~\ref{table_values} and note that $C_E$ and $D_E$ vanish)
\begin{multline}
\Braket{n}_{\text{ss}} = 1 + \sin 2\theta \cos^2 \frac{\psi}{2} \cos \phi + \cos \theta \sin \psi \cos \delta\\ + \sin \theta \sin \psi \cos(\phi-\delta).
\label{sec3.1eq29}
\end{multline}
The maximum photon number corresponds to $\delta = 0$ and $\phi = 0$. The variation of the mean number of photons with respect to the remaining parameters $\theta$ and $\psi$ is shown in Fig.~\ref{sec3.1fig1}. It is seen that the maximum number corresponds to the symmetric W state
\begin{equation}\label{eq_W}
  \Ket{W} = \left( \Ket{gge} + \Ket{geg} + \Ket{egg} \right)/\sqrt{3},
\end{equation}
which is known for its robust entanglement~\cite{nielsenbook}. This state yields $\Braket{n}_{\text{ss}}=3$, which means that coherences (correlations) in the W state increase the photon number in thermal equilibrium from the value $\langle n\rangle_{ss}^{(0)}=1$, which would be the case for a phase-averaged W state.

\par

This amplification of photon population in the cavity is here due to Dicke superradiance~\cite{dicke1954coherence}: The quantum interference in the W state enhances the processes 
described in the master equation, which equilibrate the cavity field to a canonical thermal state. Although each cluster is in a pure state, the entropy of the cavity increases via the partial-trace operation after each interaction, which removes the information about the atomic state. The crucial contribution of the heat-exchange coherences present in the W state~\eqref{eq_W} can be traced to the effective temperature $T\approx3.47\hbar\omega/{\kB}$ that these coherences induce as to compared to the temperature obtained for its phase-averaged (classically-correlated) counterpart (i.e., without any heat-exchange coherences), $T_0\approx 1.44\hbar\omega/{\kB}<T$. This temperature $T_0$ is solely determined by the populations of the computational-basis states and may hence be thought of a ``classical'' effect. By contrast, the augmented temperature $T>T_0$ stems from the heat-exchange coherences (that here lead to constructive quantum interference). The deviation from $T_0$ is thus of quantum-mechanical origin.

\par

Therefore, we conclude that the symmetric W state provides the highest equilibrium temperature to the cavity field among the entangled singly-excited states. This comes about since in the symmetric W state all contributions $H_\mathrm{int}\Ket{W}_\mathrm{gen}$ in Eq.~\eqref{eq_H_int} add up coherently, allowing for cooperatively enhanced interaction in this three-particle Dicke state~\cite{mandelbook}.

\subsubsection{\label{sec:ghzstate}GHZ states: Towards infinite effective temperature}

\begin{figure}
\centering\includegraphics[width=0.9\columnwidth]{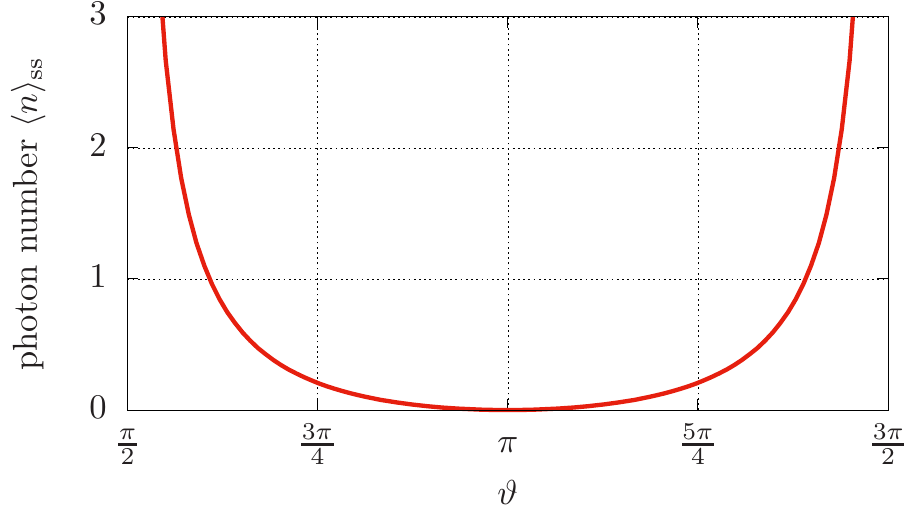}
\caption{Mean photon number~\eqref{sec3.2eq7} in a cavity pumped by atom clusters in the generalized GHZ state~\eqref{sec3.2eq6}.}
\label{GHZ_fig}
\end{figure}

Equation~\eqref{sec:single:eq7} only possesses a steady-state solution [Eq.~\eqref{sec:single:eq9}] if $a_{88}>a_{11}$ (cf. Table~\ref{table_values}). This condition can be fulfilled for a generalized GHZ state parameterized by $\vartheta$,
\begin{equation}
\Ket{\text{\text{GHZ}}}_\mathrm{gen} = \cos \frac{\vartheta}{2} \Ket{eee} + \sin \frac{\vartheta}{2} \Ket{ggg}, \label{sec3.2eq6}
\end{equation}
for which the mean photon number in thermal equilibrium becomes
\begin{equation}
\Braket{n}_{\text{ss}} = \frac{a_{11}}{a_{88}-a_{11}}=\frac{\cos^2 \frac{\vartheta}{2}}{\sin^2 \frac{\vartheta}{2} -\cos^2 \frac{\vartheta}{2}}.\label{sec3.2eq7}
\end{equation}
We have plotted the steady-state photon number as a function of $\vartheta$ in Fig.~\ref{GHZ_fig}. The figure shows that as $\vartheta\rightarrow \pi/2$ one reaches the micromaser threshold, where the mean photon number diverges, signifying an infinite effective
temperature of the bath. 

\par

Let us now consider the case $\vartheta=\pi/2$ that yields the GHZ state
\begin{equation}
  \Ket{\text{GHZ}}= \left( \Ket{ggg} + \Ket{eee} \right) /\sqrt{2}.
\end{equation}
The only nonzero coherences are then $a_{18}$ and $a_{81}$, which according to Fig.~\ref{sec2fig4} are ineffective. Indeed, the only nonvanishing parameters of the master equation~\eqref{sec:single:eq1} are $r_e=r_g=3/2$. Such parameters in the thermal Lindblad equation
\begin{equation}
\dot{\rho}(t)=\frac{3\mu}{4} \left(\mathbb{L}_e + \mathbb{L}_d \right)
 \label{sec3.2eq2}
\end{equation}
correspond, according to the KMS detailed-balance condition, to an \emph{infinite temperature} of the bath. A beam of GHZ states thus cannot thermalize a cavity mode. Indeed, the mean photon number in the cavity grows in an unbounded fashion,
\begin{equation}
  \Braket{\dot{n}}= \frac{3\mu}{2},
\end{equation}
according to Eq.~\eqref{sec:single:eq7}. 

\par

Similarly to the symmetric W state, 
a nearly symmetric GHZ-type state (cf. Fig.~\ref{fig_states}) is the optimal choice for reaching high cavity temperatures. However, the mechanism is entirely different in
the two cases. The symmetry in the W state allows for constructive quantum interference (superradiance), so that the enhancement
is purely quantum-mechanical. By contrast, the nearly-symmetric entangled GHZ state allows us to approach the maser threshold. The photon number~\eqref{sec3.2eq7} only depends on the populations since, according to Fig.~\ref{sec2fig4}, all coherences in the state~\eqref{sec3.2eq6} are ineffective. Hence, the phase-averaged counterpart of Eq.~\eqref{sec3.2eq6} results in the same effective temperature as that of the state~\eqref{sec3.2eq6}, i.e., $T_0=T$.

\subsubsection{States leading to ultrahigh temperatures of the cavity field}

Let us \emph{incoherently} mix W states with states that we denote as E states that belong to the general class of three-atom W states with two excitations in the upper blue triangle in Fig.~\ref{sec2fig3} (which are also Dicke states~\cite{agarwalQO}),
\begin{equation}
\Ket{\text{E}} = \frac{1}{\sqrt{3}} \left(\Ket{eeg} + \Ket{ege} + \Ket{gee} \right). 
\label{secE:eq1}
\end{equation}
According to Table~\ref{table_values} they contribute to $C_E$ and $D_E$ and correspond to $r_e = 4$ and $r_g=3$, namely their rate of absorption surpasses the emission rate, leading [by the KMS detailed balance condition~\eqref{eq_kms}] to a \emph{negative temperature}, which is outside the scope of this paper.

\par

The chosen nearly equal mixture of W- and E-states has the form 
\begin{equation}\label{eq_we}
\rho_{\text{WE}}=\left(\frac{1}{2}+\varepsilon\right)\Ket{W}\!\Bra{W}
+\left(\frac{1}{2}-\varepsilon\right)\Ket{E}\!\Bra{E},
\end{equation}
where $0<\varepsilon\ll 1$ is a small positive number. This state is a mixture of the two coherence triangles indicated 
by solid blue lines in Fig.~\ref{sec2fig3}. The corresponding nonvanishing parameters of the master equation~\eqref{sec:single:eq1} $r_e = 7/2-\varepsilon$ and $r_g = 7/2+\varepsilon$ (cf. Table~\ref{table_coefficients}) imply that $r_g>r_e$, so that this mixed state corresponds to a \emph{positive and finite} effective temperature, $T\approx 7\varepsilon\hbar\omega/4{\kB}$. The corresponding classical-like phase-averaged counterpart of Eq.~\eqref{eq_we} would
thermalize the cavity to $T_0=3\varepsilon\hbar\omega/4{\kB}$. Here, two enhancement factors are involved. The first factor is $C$ (cf. Table~\ref{table_values}), enhancing
$T_0$ to higher temperatures by
the quantum interferences due to the coherences in the $W$ and $E$ states. The second factor is the classical enhancement of $T_0$
due to the operation near the maser threshold.

\section{\label{sec:conclusions}Discussion}

We have studied the thermodynamic implications of a generalized micromaser model wherein the cavity mode interacts with a beam of quantum-coherent or quantum-correlated multiatom clusters. Our central goal has been to classify the states of such clusters prior to their injection into the cavity according to their ability to fuel the cavity field as ``working fluid'' in a machine of either the first kind (thermo-mechanical engine) or the second kind (heat engine). To this end we have derived a Lindblad master equation for the cavity field mode that describes absorption- and emission of the field, its coherent displacement and squeezing caused by the atoms that may act, respectively, as a thermal, displaced-thermal or squeezed-thermal bath. These distinct Gaussian processes that the field may undergo are determined by the prefactors of the respective terms in the master equation that are, in turn, determined by disjoint blocks (coherences) of the multiatom density matrix.

\par

The main results of our analysis are as follows:
\begin{itemize}
\item An important insight that we have obtained is that two- and three-atom clusters are capable of acting as fuel for both kinds of machines in a highly effective fashion, so that there is no need to involve larger clusters. Still, a larger number of coherences as the cluster grows in size may further enhance the work output.
\item For machines of the first kind, our analysis has revealed a particularly promising, simple, fuel in the form of two-atom clusters whose state is a nearly equal superposition of doubly-excited and doubly-ground states. Such a state is expected to give rise to very large squeezing of the cavity field. It may thus present a far superior alternative to existing squeezing schemes of cavity fields~\cite{wenger2004nongaussian,su2007experimental,eberle2010quantum}. Such a strong squeezing may have fascinating applications~\cite{eberle2010quantum} also outside of quantum thermodynamics. Our interest here is that this strong squeezing source may fuel a cavity field in a hybrid thermo-mechanical machine~\cite{niedenzu2015efficiency} with nearly 100\% efficiency, at the expense of mechanical work supplied by the two-atom clusters.
\item For machines of the second kind, we have found W states of three-atom clusters to act as conventional heat-baths fuel at a positive finite temperature that is controllable by the W state. By contrast, three-atom GHZ- and E-states have been found to correspond to effective baths at infinite or negative temperatures, respectively, that do not allow for a thermal steady-state solution for the cavity field. On the other hand, nearly-equal mixtures of W and E states have been identified as fuel capable of thermalizing the cavity field to an \emph{ultrahigh temperature}.
\end{itemize}

\par 

To conclude, our results are potentially useful for the design of thermal and nonthermal machines based on micromaser setups. The availability of all Gaussian processes via preparation of two- and three-atom clusters allows to implement heat engines (wherein the cavity field is thermalized) but also thermo-mechanical engines (wherein the cavity mode is coherently displaced or squeezed).

\par

We wish to stress the feasibility of the diverse forms of state preparation of multiatom clusters (prior to their injection into the cavity) we have employed in our analysis:
\begin{itemize}
\item The arsenal of quantum gate operations~\cite{nielsenbook} can in principle prepare two or three trapped atoms in an entangled state on demand, but such preparation may require single-atom addressability.
\item Alternatively, W states can be generated via quantum feedback control~\cite{huang2013generation} or at fusion-based light--matter interfaces~\cite{zang2015generating}. Multipartite entangled states may also be generated via photon-mediated interactions, as recently discussed in~\cite{aron2016photon}.
\item Another alternative is an optimized probabilistic scheme for multiatom entangled-state preparation in a cavity~\cite{mandilara2007control}.
\item For two-atom entangled-state preparation we may resort to controlled diatomic dissociation~\cite{kurizki1987theory}, collisions in a cavity~\cite{deb1999formation} or long-range dipole--dipole interactions~\cite{shahmoon2013nonradiative}.
\end{itemize}

\par

On the fundamental side, our results provide clues to the thermalization or nonthermalization of a system (here the cavity field) via its contact with quantum-correlated multipartite clusters that act as nonthermal baths. Such processes reflect the subtle rapport between quantum correlations in the bath and thermalization~\cite{PRE89}.
\vspace{6pt}
\begin{acknowledgments}
  C.~B.~D. is thankful for intuitive discussions with A. Levent Subas{\i} and A. \"{U}mit Hardal. \"{O}.~E.~M. acknowledges support from Ko\c{c} University and Lockheed Martin Corporation University Research Agreement. G. K. acknowledges the ISF and BSF for support.
\end{acknowledgments}

\section*{Author Contributions}\phantomsection
\"O.~E.~M. conceived the idea. C.~B.~D., \"O.~E.~M., and W.~N. equally contributed to the analytical and 
numerical calculations. All authors revised and wrote the paper in collaboration. 
All authors have read and approved the final manuscript.

\appendix

\begin{widetext}

\section{Time-evolution operator for a one-atom micromaser}\label{app_one_atom}

For one atom the time-evolution operator
\begin{equation}
  U(\tau)= \exp(-i H_{\text{\text{int}}} \tau)
\end{equation}
to second order in $g\tau$ readily evaluates to
\begin{equation}\label{eq_app_Utau_one}
  U(\tau)\approx\mathds{1}-ig\tau 
  \begin{pmatrix}
    0&a\\
    a^\dagger&0
  \end{pmatrix}
  -\frac{(g\tau)^2}{2}
  \begin{pmatrix}
    0&a\\
    a^\dagger&0
  \end{pmatrix}^2
  =
  \begin{pmatrix}
    1-\frac{1}{2}(g\tau)^2(a^\dagger a+1) & -ig\tau a\\
    -ig\tau a^\dagger & 1-\frac{1}{2}(g\tau)^2a^\dagger a
  \end{pmatrix}
  .
\end{equation}

\section{Time-evolution operator for a two-atom micromaser}\label{app_two_atoms}

The time-evolution operator
\begin{equation}
  U(\tau)= \exp(-i H_{\text{\text{int}}} \tau) = \exp(-ig\tau  P) \label{secA.eq1}
\end{equation}
of the joint cavity--atoms system can be computed to second order in $g\tau$ using the collective angular-momentum operators $S_{\pm}=\sum_{j=1}^2\sigma_j^\pm$ such that
\begin{equation}
  P=aS_++a^\dagger S_-.  
\end{equation}
The latter can be decomposed into irreducible subspaces by changing from the computational basis (spanned by products of single-atom $\Ket{e}$ and $\Ket{g}$ states, cf. Fig.~\ref{sec2fig2}) to the basis of Dicke states~\cite{mandelbook} by means of the transformation matrix
\begin{equation}
T=\left(
  \begin{array}{cccc}
    1 & 0 & 0 & 0 \\
    0 & \frac{1}{\sqrt{2}} & 0 & \frac{1}{\sqrt{2}} \\
    0 & \frac{1}{\sqrt{2}} & 0 & -\frac{1}{\sqrt{2}} \\
    0 & 0 & 1 & 0 \\
  \end{array}
\right).
\end{equation}
The addition of two spin-$1/2$ gives rise to a triplet and a singlet, $\frac{1}{2} \otimes \frac{1}{2} = 1 \oplus 0$. As a consequence, 
\begin{equation}
T^{\dagger}PT =
\begin{pmatrix}
  0 & \sqrt{2}a & 0 & 0 \\
  \sqrt{2}a^\dagger & 0 & \sqrt{2} a & 0 \\
  0 & \sqrt{2}a^\dagger & 0 & 0 \\
  0 & 0 & 0 & 0
\end{pmatrix}=
\left( \begin{array}{c c c}
P_{1} & \\
 & P_{0}
\end{array} \right)
\end{equation}
and the propagator is given by the direct sum
\begin{equation}
  U(\tau)=U_{1}(\tau)\oplus U_{0}(\tau),
\end{equation}
where to second order in $g\tau$
\begin{equation}
  U_k(\tau)\approx\mathds{1}_k-ig\tau P_k-\frac{(g\tau)^2}{2}P_k^2.
\end{equation}
Here $\mathds{1}_k$ denotes the unit matrix of the same dimensionality as $P_k$. Explicitly, we find
\begin{equation}
  U_{1}(\tau)=\left(
      \begin{array}{ccc}
        1-(g\tau)^2 (a^\dagger a+1) & -i \sqrt{2} a g\tau & -a^2 (g\tau)^2 \\
        -i \sqrt{2} {a^\dagger} g\tau & 1-(g\tau)^2 (2 a^\dagger a+1) & -i \sqrt{2} a g\tau \\
        -{a^\dagger}^2 (g\tau)^2 & -i \sqrt{2} {a^\dagger} g\tau & 1-(g\tau)^2 a^\dagger a \\
      \end{array}
    \right)
\end{equation}
and
\begin{equation}
  U_{0}(\tau)=1.
\end{equation}

\par

Transforming back to the computational basis $\{\Ket{ee},\Ket{eg},\Ket{ge},\Ket{gg}\}$ yields
\begin{equation}\label{eq_app_Utau_two}
  U(\tau)=\left(
    \begin{array}{cccc}
      1-(g\tau)^2 (a^\dagger a+1) & -i a g\tau & -i a g\tau & -a^2 (g\tau)^2 \\
      -i {a^\dagger} g\tau & 1-\frac{1}{2} (g\tau)^2 (2 a^\dagger a+1) & -\frac{1}{2} (g\tau)^2 (2 a^\dagger a+1) & -i a g\tau \\
      -i {a^\dagger} g\tau & -\frac{1}{2} (g\tau)^2 (2 a^\dagger a+1) & 1-\frac{1}{2} (g\tau)^2 (2 a^\dagger a+1) & -i a g\tau \\
      -{a^\dagger}^2 (g\tau)^2 & -i {a^\dagger} g\tau & -i {a^\dagger} g\tau & 1-(g\tau)^2 a^\dagger a \\
    \end{array}
  \right).
\end{equation}

\section{Time-evolution operator for a three-atom micromaser}\label{app_three_atoms}
For three particles one proceeds exactly like in the preceding section. The transformation matrix now reads~\cite{mandelbook}
\begin{equation}
T = \left(
\begin{array}{cccccccc}
 1 & 0 & 0 & 0 & 0 & 0 & 0 & 0 \\
 0 & \frac{1}{\sqrt{3}} & 0 & 0 & 0 & 0 & -\sqrt{\frac{2}{3}} & 0 \\
 0 & \frac{1}{\sqrt{3}} & 0 & 0 & -\frac{1}{\sqrt{2}} & 0 & \frac{1}{\sqrt{6}} & 0 \\
 0 & \frac{1}{\sqrt{3}} & 0 & 0 & \frac{1}{\sqrt{2}} & 0 & \frac{1}{\sqrt{6}} & 0 \\
 0 & 0 & \frac{1}{\sqrt{3}} & 0 & 0 & -\frac{1}{\sqrt{2}} & 0 & -\frac{1}{\sqrt{6}} \\
 0 & 0 & \frac{1}{\sqrt{3}} & 0 & 0 & \frac{1}{\sqrt{2}} & 0 & -\frac{1}{\sqrt{6}} \\
 0 & 0 & \frac{1}{\sqrt{3}} & 0 & 0 & 0 & 0 & \sqrt{\frac{2}{3}} \\
 0 & 0 & 0 & 1 & 0 & 0 & 0 & 0 \\
\end{array}
\right).
\label{secA.eq4}
\end{equation}
The addition of three spin-$1/2$ gives rise to a quadruplet and two doublets, $\frac{1}{2} \otimes \frac{1}{2} \otimes \frac{1}{2} = \frac{3}{2} \oplus \frac{1}{2} \oplus \frac{1}{2}$. As a consequence, 
\begin{equation}
T^{\dagger}PT =
\left(
\begin{array}{cccccccc}
 0 & \sqrt{3} a & 0 & 0 & 0 & 0 & 0 & 0 \\
 \sqrt{3} a^\dagger & 0 & 2 a & 0 & 0 & 0 & 0 & 0 \\
 0 & 2 a^\dagger & 0 & \sqrt{3} a & 0 & 0 & 0 & 0 \\
 0 & 0 & \sqrt{3} a^\dagger & 0 & 0 & 0 & 0 & 0 \\
 0 & 0 & 0 & 0 & 0 & a & 0 & 0 \\
 0 & 0 & 0 & 0 & a^\dagger & 0 & 0 & 0 \\
 0 & 0 & 0 & 0 & 0 & 0 & 0 & a \\
 0 & 0 & 0 & 0 & 0 & 0 & a^\dagger & 0 \\
\end{array}
\right)=
\left( \begin{array}{c c c}
P_{3/2} & & \\
 & P_{1/2} & \\
 & & P_{1/2}
\end{array} \right)
\label{secA.eq3}
\end{equation}
and the propagator is given by the direct sum
\begin{equation}
  U(\tau)=U_{3/2}(\tau)\oplus U_{1/2}(\tau)\oplus U_{1/2}(\tau),
\end{equation}
where
\begin{equation}
  U_{3/2}(\tau)=\left(
    \begin{array}{cccc}
      1-\frac{3}{2} (g\tau)^2 (a^\dagger a+1) & -i \sqrt{3} a g\tau & -\sqrt{3} a^2 (g\tau)^2 & 0 \\
      -i \sqrt{3} a^\dagger g\tau & 1-\frac{1}{2} (g\tau)^2 (7 a^\dagger a+4) & -2 i a g\tau & -\sqrt{3} a^2 (g\tau)^2 \\
      -\sqrt{3} {a^\dagger}^2 (g\tau)^2 & -2 i a^\dagger g\tau & 1-\frac{1}{2} (g\tau)^2 (7 a^\dagger a+3) & -i \sqrt{3} a g\tau \\
      0 & -\sqrt{3} {a^\dagger}^2 (g\tau)^2 & -i \sqrt{3} a^\dagger g\tau & 1-\frac{3}{2} (g\tau)^2 a^\dagger a \\
    \end{array}
  \right)
\end{equation}
and
\begin{equation}
  U_{1/2}(\tau)=\left(
    \begin{array}{cc}
      1-\frac{1}{2} (g\tau)^2 (a^\dagger a+1) & -i a g\tau \\
      -i a^\dagger g\tau & 1-\frac{(g\tau)^2 a^\dagger a}{2} \\
    \end{array}
  \right).
\end{equation}

\par

Transforming back to the computational basis (cf. Fig.~\ref{sec2fig2}) yields the matrix elements
\begin{eqnarray}\label{eq_app_Utau_three}
\begin{split}
&U_{11} = \frac{1}{2} \left(2-3 (g\tau  )^2 (a^{\dagger} a+1) \right), \\
&U_{21} = U_{31} = U_{52} = U_{62} = U_{53} = U_{73} = U_{41} \\
&\hspace{5.5mm}= U_{85} = U_{64} = U_{74} = U_{86} = U_{87} = - i g\tau   a^{\dagger},\\
&U_{12} = U_{13} = U_{25} = U_{26} =  U_{35} =  U_{37} = U_{14} \\
&\hspace{5.5mm}=  U_{58} = U_{46} = U_{47} = U_{68} = U_{78} = -i a g\tau \\
&U_{51} = U_{61} = U_{71} = U_{82} = U_{83} = U_{84} = -(g\tau  )^2\left(a^{\dagger }\right)^2,\\
&U_{15} = U_{16} = U_{17} = U_{28} = U_{38} = U_{48} =  - (g\tau  )^2 a^2 ,\\
&U_{22} = U_{33} = U_{44} = 1-\frac{1}{2} (g\tau  )^2 (3a^{\dagger} a+2),\\
&U_{32} = U_{42} = U_{23} = U_{43} = U_{65} = U_{75} = U_{24} = U_{34}  \\
&\hspace{5.5mm}= U_{56} = U_{76} = U_{57} = U_{67} = -\frac{1}{2} (g\tau  )^2 (2a^{\dagger} a+1),\\
&U_{55} = U_{66} = U_{77} = 1-\frac{1}{2} (g\tau  )^2 (3 a^{\dagger} a+1),\\
&U_{88} = 1-\frac{3}{2} (g\tau  )^2 a^{\dagger} a
\end{split}
\end{eqnarray}
of the time-evolution operator. The remaining elements evaluate to zero.

\end{widetext}

\section{Maser threshold}\label{app_maser_threshold}

It is illuminating to derive the threshold condition from the mean photon number, following Ref.~\cite{PRE89}. We first
take $\lambda=\xi=0$ and thereby eliminate the coherent-displacement and squeezing terms. 
According to Eq.~(\ref{sec2eq4}), the cavity density matrix will change to
\begin{equation}
\rho(t_j + \tau) \approx  (g\tau)^2 \left( \frac{r_e}{2} \mathbb{L}_e +
 \frac{r_g}{2}  \mathbb{L}_d \right) + \rho(t_j)
\label{secCC:eq3.2}
\end{equation}
after the passage of the $j$th atom during the short interaction time $\tau$.
The mean photon number can be calculated to be
\begin{equation}
\Braket{\hat{n} (t_j + \tau)} = \text{Tr} \left[ \rho(t_j + \tau) \hat{n} \right] = k \Braket{\hat{n}(t_j)} + (g\tau)^2 \frac{r_e}{2} .
\label{secCC:eq3.3}
\end{equation}
The change of the mean photon number between consecutive injections of two atom clusters is determined by
the increment ratio $k$, which is given by
\begin{equation}
k = 1 - (g\tau  )^2 \frac{r_g-r_e}{2}. 
\label{secCC:eq3.5}
\end{equation}
Assuming that the cavity is initially in the vacuum state, the last term in 
Eq.~(\ref{secCC:eq3.3}) yields the mean number of photons after the first-cluster passage, $\Braket{\hat{n}(\tau)} = (g\tau)^2r_e/2$.
After the $j$th cluster passage the mean photon number rises to
\begin{equation}
\Braket{\hat{n}(t_j)} = \sum_{i=1}^j k^{i-1} \Braket{\hat{n}(\tau)}. 
\label{secCC:eq3.7}
\end{equation}
The summation in Eq.~(\ref{secCC:eq3.7}) is convergent if $k < 1$, which is equivalent to the threshold condition $r_g>r_e$. As $j\rightarrow\infty$, the summation converges to Eq.~\eqref{sec:single:eq9}.

\end{document}